\documentclass[pra,showpacks, showkeys, reprint,amsmath,amssymb,aps]{revtex4-2}
\usepackage{xcolor}
\usepackage{graphicx,subcaption}% Include figure files
\usepackage{anyfontsize}
\usepackage{dcolumn}% Align table columns on decimal point
\usepackage{bm}% bold math
\usepackage{amsthm}
\usepackage{epstopdf}
\usepackage{nccmath}
\usepackage{float}
\usepackage{multirow}
\newtheorem{theorem}{Theorem}

\usepackage{appendix}
\usepackage{balance}
\usepackage{silence}
\usepackage{mathtools}
\newlength{\arrow}
\newcommand{\bra}[1]{\langle#1|} % Bra
\newcommand{\ket}[1]{|#1\rangle} % Ket
\newcommand{\braket}[2]{ \langle #1 | #2 \rangle} %BraKet

\begin{document}
	
	\preprint{APS/123-QED}
	
	\title{Condition for the generation of the secret key in a BB84 like quantum key distribution protocol}
	
	\author{Rashi Jain}
	\email[]{rashijain\_23phdam07@dtu.ac.in}
	\author{Satyabrata Adhikari}
	\email[]{satyabrata@dtu.ac.in}
	\affiliation{Department of Applied Mathematics,\\ Delhi Technological University, Delhi-110042, India}
	%\date{\today}
	%
	\begin{abstract}
		
		Woodhead [Phys. Rev. A \textbf{88}, 012331 (2013)] derived the lower bound of the secret key rate for a  Bennett-Brassard (BB84) like quantum key distribution protocol under collective attacks. However, this lower bound does not always assure the generation of the secret key and thus the protocol may have to be aborted sometimes. Thus, we modify the Woodhead's lower bound of the secret key rate in such a way that the secret key is always generated in a BB84 like quantum key distribution protocol. We show the non-linear relationship between the lower bound of the secret key rate with the error rate and fidelity. Exploiting the obtained modified lower bound of the secret key rate, we  analyze two state dependent quantum cloning machines such as (i) Wootters-Zurek QCM and (ii) Modified Buzek-Hillery QCM constructed by fixing the cloning machine parameters of Buzek Hillery quantum cloning machine (QCM), which may be used by the eavesdropper to extract information from the intercepted state. We, thereafter, show that it is possible for the communicating parties to distill a secret key, even in the presence of an eavesdropper. Moreover, we also discuss the effect of the efficiency of the QCM on the generation of the secret key for a successful key distribution protocol. 
	\end{abstract}
	
	%\pacs{98.80.-k, 98.80.Es}
	%Use showkeys class option if keyword
	\maketitle{}
	
	%\tableofcontents
	\section{Introduction}
	One of the most challenging problems in the modern world is to communicate securely, over an insecure channel \cite{merkle_1978}. Quantum cryptography plays a vital role in providing security in the areas related to quantum technologies \cite{gisin_2002}. Quantum key distribution (QKD) is one of the most advanced applications of quantum cryptography. In a QKD protocol, a secure key is established between two or more trusted communicating parties following the laws of quantum mechanics \cite{ballentine_2014}. This key is then used for secure communication over a quantum channel. Various QKD protocols have been proposed since early 1980s \cite{bb84,jain_2024}. The security of the key being distributed is the utmost priority for the communicating parties \cite{scarani_2009}. The distribution of the secret key is a challenging task as there might be some external noise or an eavesdropper present in the quantum channel, which may tamper with the key \cite{tan_2025}. \\
	There are broadly two classes of QKD implementation namely, prepare-and-measure QKD and, entanglement-based approach \cite{ekert_1991}. In this paper, we will focus on the prepare-and-measure QKD scheme \cite{fung_2006}. In this scheme, the sender, say Alice, prepares and transmits quantum states randomly, on one of the previously decided bases to the receiver, say Bob. Upon receiving the states, Bob performs quantum measurements upon them on any one of the basis in a random manner. The state preparation basis and state measurement basis are then discussed in the public channel to assure the correctness of the key. This forms a \textit{raw key} \cite{young_2026}. Alice and Bob then chooses randomly any subset of the raw key, discuss their prepared states and calculate the error occurred in the selected subset. If the error is too high, they abort the protocol, else they continue \cite{metger_2023}. The remaining qubits form a \textit{sifted key}, which can be improved by applying error correction schemes \cite{dixon_2014,pastushenko_2023} and privacy amplification schemes \cite{horvath_2011,tang_2019}. In an ideal case, Bob's string is identical to Alice's string, and the key rate is considered to have no dark counts and perfect detector efficiency \cite{pirandola_2020}. Although, in realistic situations, this might not be the case. An eavesdropper, say Eve, might be present in between, who is in general assumed to posses every possible power that is compatible within the laws of quantum mechanics \cite{barrett_2005}. The most general form of attacks considered in QKD are joint attacks. An important class of joint attacks is the collective attack \cite{biham_2002}. Eve may intercept Alice's state before reaching to Bob in the midway and she may apply an unitary transformation on it in order to know about the secret key. Since the above procedure is repeating several times so after the classical communication has been completed between Alice and Bob, Eve separately applies the same operation to each state that she intercepted from Alice \cite{acin_2007}. This is known as a \textit{collective attack} \cite{renner_thesis}.\\
	The interference of Eve generate some errors in the key distribution scheme. This raises the problem of determining the number of key bits that can be securely extracted from a QKD protocol, or in other words, how much error can be tolerated by the QKD scheme such that Alice and Bob are still able to distill a key \cite{diamanti_2016}. In this regard, in 2004, Gottesman et al, proposed the achievable key rate for the BB84 protocol \cite{bb84}, which is given by \cite{gottesman_2004}
	\begin{align}
		r=1-H_2(\delta)-H_2(\delta_p)
	\end{align}
	where $H_2(\delta)$ denotes the Shannon entropy of a random variable $\delta$, representing the fraction of the sifted key bits being sacrificed to perform error correction, and $H_2(\delta_p)$ represents the fraction of the sifted key bits being sacrificed to perform privacy amplification. The achievable key rate is a non-linear function of $\delta$ and $\delta_p$ \cite{liu_2025,shan_2024,feng_2025,wang_2026,gao1_2024,gao1_2025,gao2_2024,gao2_2025}. Later, in 2005, Devetak and Winter proposed a lower bound on the asymptotic key rate that is secure against an adversary under collective attacks \cite{devetak_2005}. Kraus et al \cite{kraus_2005} proposed lower and upper bounds on the secret key using one way classical communication QKD protocols. Till date there have been many theories in the literature for the upper bound of key rate in various kinds of QKD protocols \cite{koashi_2009, curty_2009,maroy_2010, kaur_2022}, but only a few works comprises of lower bound of secret key rates \cite{kraus_2005, zhang_2017}. Therefore, in this paper, we derive the possible lower bound, for a successful BB84 like protocol. We call BB84 like protocol is successful, if the key rate is positive. We further find the upper bound on the error rate such that the two communicating parties are able to distill a secret key, even in the presence of Eve, where she uses a quantum cloning transformation to extract information from the intercepted state from the sender, Alice. \\
	The aim of this paper is three-fold. Firstly, we observe that the lower bound for a key rate introduced by \cite{woodhead_2013} is based on Eve's fidelity and the errors introduced in the system for a BB84-like protocol. While this bound provides insights about the security related to key generation, it fails to analyze the successful completion of the protocol in terms of positive secret key generation. In particular, it does not highlight the parameters under which a non-zero secret key can be distilled between the communicating parties. Thus, our aim is to find the estimate of the lower bound possible on the secret key rate such that the protocol is completed successfully, i.e., the key is formed positively. Secondly, our aim is to find whether Eve's cloning transformation has some effect on the key rate, or the introduced errors. To study this, we find the maximum possible upper bound of the introduced errors such that the communicating parties are able to distill a secret key. Thirdly, we aim to find if there exists an upper bound on the efficiency of the QCM that may effect the generation of the secret key. Our work addresses a key limitation of the key rate bound derived by Woodhead. The limitation is that the lower bound derived by Woodhead does not assure the	success of the protocol, i.e., successful key generation between the communicating parties. This means that the	protocol may be aborted if the key rate is found to be negative, which restricts practical applicability. In contrast, we modify this lower bound to guarantee a strictly positive key rate in BB84-like protocol, even in the presence of imperfections, such as a potential eavesdropper. Further, we find an upper bound on the error rate such that the communicating parties are still be able to distill a key even in the presence of an eavesdropper.\\
	Moreover, Woodhead demonstrated the cloning bound using the Wootters-Zurek cloning machine, whereas	we explicitly analyzed various state-dependent quantum cloning attacks, such as the Wootters-Zurek and the	modified Buzek-Hillery QCM, thereby addressing more realistic eavesdropping strategies. Thus, in our work, we	extend the original theoretical contribution mathematically by making it more applicable to real-world cryptographic schemes.\\
	%In the previous works, we do not find any such relationship. Thus, our work explicitly highlights the relationship between the efficiency of the QCM that may be used by the eavesdropper and the successful secret key generation in a BB84-like QKD protocol.}\\
The rest of the paper is organized as follows: In Section II, we will discuss the BB84 like QKD protocol and provide the theoretical proposal for the generation of Alice's state in a laboratory, which she uses for QKD. We also clearly define the role of the eavesdropper in the QKD protocol. In Section III, we provide the estimate of the lower bound for a positive key rate. Moreover, for a given fidelity between the states resulting from the output of the QCM, we find the upper bound on the error rate upto which the protocol proves to be successful. In Section IV, we study the state dependent symmetric QCMs, such as the Wootter-Zurek QCM, and the modified Buzek-Hillery QCM. We find the upper bound on error rate, for a given fidelity and analyze the value of secret key rate, such that the protocol is successful. In Section V, we study the effect of efficiency of the QCM that may be used by Eve in a successful QKD protocol. Finally, we conclude with a brief summary in Section VI.

\section{BB84 like Protocol}
Consider the following setting of BB84 like protocol, which is similar to the BB84 protocol: Two communicating parties, namely, Alice and Bob are connected via an untrusted quantum channel and they wish to establish a secure secret key for further communication. To start with, let us suppose that Alice possesses an unbiased source that produces the two-qubit state of the following form
\begin{align}
	\begin{split}	\ket{\Phi}_{AA}&=\ket{0}_{A}\otimes\ket{S}_A+\ket{1}_A\otimes \ket{S'}_A
		\label{alice_two_qubit_state}
	\end{split}
\end{align}
where $\ket{S}\equiv \ket{\phi}~\text{or}~ \ket{\psi}$ and $\ket{S'}\equiv \ket{\phi'}~\text{or}~ \ket{\psi'}$.
These four quantum states, either from the \textquotedblleft$Z$ basis\textquotedblright~or  \textquotedblleft$X$ basis\textquotedblright~ (these bases may not be identical to the actual $Z$ and $X$ basis known in quantum mechanics) at random, with equal probability \cite{maroy_2010}. The relationship between the usual $Z$ and $X$ basis and new $Z'$ and $X'$ basis is given in Appendix A. Let us further consider that the two orthogonal states in \textquotedblleft$Z$ basis\textquotedblright~are $\{\ket{\phi},\ket{\phi'}\}$, which is of the form
\begin{equation}
	\begin{split}
		\ket{\phi}=\alpha \ket{0}+\beta \ket{1}\\
		\ket{\phi'}=\alpha \ket{1}-\beta\ket{0}
	\end{split}
	\label{z_basis_states}
\end{equation}
where the real numbers $\alpha$ and $\beta$ satisfy the normalization condition $\alpha^2 +\beta^2=1$. The orthogonal states of the \textquotedblleft$X$ basis\textquotedblright~can be given by $\{\ket{\psi},\ket{\psi'}\}$. The density operators corresponding to these states can be given by $\rho=\ket{\phi}\bra{\phi}$, $\rho'=\ket{\phi'}\bra{\phi'}$, $\sigma=\ket{\psi}\bra{\psi}$ and $\sigma'=\ket{\psi'}\bra{\psi'}$.
Alice performs measurement in the computational basis $\{\ket{0},\ket{1}\}$ and according to the measurement outcome, she may get one of the four states $\{\ket{\phi},\ket{\phi'},\ket{\psi},\ket{\psi'}\}$. Alice then  sends the resulting single qubit state obtained from the measurement, to Bob. Bob chooses either \textquotedblleft$Z$ basis\textquotedblright~or \textquotedblleft$X$ basis\textquotedblright~at random and performs measurement on the received state, and keeps the sequence obtained in terms of bits 0 and 1 respectively. Thereafter, Alice and Bob discuss their state preparation basis and state measurement basis, respectively, through the public channel. This forms a raw key between Alice and Bob. We note here that in the variant of BB84 protocol, which is called as the BB84 like protocol, the sender Alice and the receiver Bob use only the \textquotedblleft$Z$ basis\textquotedblright~results for the secret key generation.  Then they select a subset of this raw key to calculate the \textquotedblleft$Z$ basis\textquotedblright~bit error rate denoted by $\delta_z$. They use the remaining data to extract the keys.\\

\subsection{Generation of Alice's two-qubit state in the laboratory }
In this section, we will show the procedure of the experimental generation of the two-qubit state possessed by Alice. Let us start with a pseudopure state (PPS), given by $\ket{00}$, which is experimentally generated by the process adopted by \cite{shruti_2015}. The two-qubit state possessed by Alice can be constructed by applying a sequence of various quantum gates upon the PPS state. The two-qubit state possessed by Alice is given by
\begin{align}
	\begin{split}	\ket{\Phi}_{AA}&=\ket{0}_{A}\otimes\ket{\phi}_A+\ket{1}_A\otimes \ket{\phi'}_A\\
		&=\frac{\alpha}{\sqrt{2}}(\ket{00}_{AA}+\ket{11}_{AA})+\frac{\beta}{\sqrt{2}}(\ket{01}_{AA}-\ket{10}_{AA})
		\label{alice_two_qubit_state_nmr}
	\end{split}
\end{align}
where $\ket{0}_A$ and $\ket{1}_A$ are the states of the classical registers possessed by Alice and $\ket{\phi}_A=\alpha\ket{0}_A+\beta\ket{1}_A$ and $\ket{\phi'}_A=\alpha \ket{1}_A-\beta \ket{0}_A$ denotes the $Z$ basis states emitted by Alice, intended for Bob.\\
The quantum circuits for the state construction consists of single-qubit rotation gate, controlled-NOT gates, controlled rotation (CROT) gate, followed by a single-qubit Hadamard gate. The unitary transformations applied are such that the normalization of the output state is preserved.\\
The single-qubit Hadamard gate for the basis $\{\ket{0},\ket{1}\}$ is defined as
\begin{align}
	H=\frac{1}{\sqrt{2}}\begin{bmatrix}
		1&1\\
		1&-1
	\end{bmatrix}
\end{align}
For the computational basis $\{\ket{00}, \ket{01}, \ket{10}, \ket{11}\}$, the ideal CNOT gate flips the target qubit (second qubit), when the control qubit (first qubit) is $\ket{1}$, which is given by
\begin{align}
	\text{CNOT}_{12}=
	\begin{bmatrix}
		1 &0 &0& 0\\
		0& 1& 0 &0\\
		0& 0& 0&1\\
		0 &0 & 1 &0
	\end{bmatrix}
\end{align}
Similarly, a controlled rotation gate (CROT-gate) is given by
\begin{align}
	\text{CROT}^\theta_{12}=\begin{bmatrix}
		1 &0 &0& 0\\
		0& 1& 0 &0\\
		0& 0& \cos \frac{\theta}{2} &-\sin \frac{\theta}{2}\\
		0 &0 & \sin \frac{\theta}{2} &\cos \frac{\theta}{2}
	\end{bmatrix}
\end{align}
where $\theta$ represents the arbitrary angle of rotation.\\
The experimental construction of the two-qubit state possessed by Alice is given by
%   {\small
	\begin{small}	
		\begin{align}
			\begin{split}
				\ket{00}\xrightarrow{\tiny{\makebox[0.8cm]{U$_1^{2\theta_1}$}}}& \cos \theta_1 \ket{00}+\sin \theta_1 \ket{10}\\
				\xrightarrow{\tiny{\makebox[0.8cm]{CNOT$_{12}$}}}& \cos \theta_1 \ket{00}+\sin \theta_1 \ket{11}\\
				\xrightarrow{\tiny{\makebox[0.8cm]{CROT$_{12}^{2\theta_2}$}}}& \cos \theta_1 \ket{00}+\sin \theta_1 \cos \theta_2 \ket{11}+ \sin \theta_1 \sin \theta_2 \ket{10}\\
				\xrightarrow{\tiny{\makebox[0.8cm]{H$_1$}}}& \frac{1}{\sqrt{2}} \bigg[(\cos \theta_1+\sin  \theta_1\sin \theta_2)\ket{00} +(\sin\theta_1\cos \theta_2)\ket{01} \\
				&+ (\cos\theta_1 -\sin\theta_1 \sin \theta_2) \ket{10}+(-\sin\theta_1\cos\theta_2) \ket{11} \bigg]\\
				\xrightarrow{\tiny{\makebox[0.8cm]{CNOT$_{12}$}}}& \frac{1}{\sqrt{2}} \bigg[(\cos \theta_1+\sin  \theta_1\sin \theta_2)\ket{00} +(\sin\theta_1\cos \theta_2)\ket{01}\\
				&+(-\sin\theta_1\cos\theta_2) \ket{10}+(\cos\theta_1 -\sin\theta_1 \sin \theta_2) \ket{11}\bigg]
			\end{split}
			\label{nmr_output}
		\end{align}
	\end{small}
	% }%

The single-qubit unitary transformation $U_{1}^{2\theta_1}$ is a transformation that rotates the first qubit by arbitrary angle $\theta_1\in (0,\pi/2)$. The controlled-NOT operation is applied by the implementation of $\text{CNOT}_{12}$, considering the first qubit as the control qubit and second qubit as the target qubit.
\begin{figure}[H]
	\centering
	\includegraphics[width=1\linewidth]{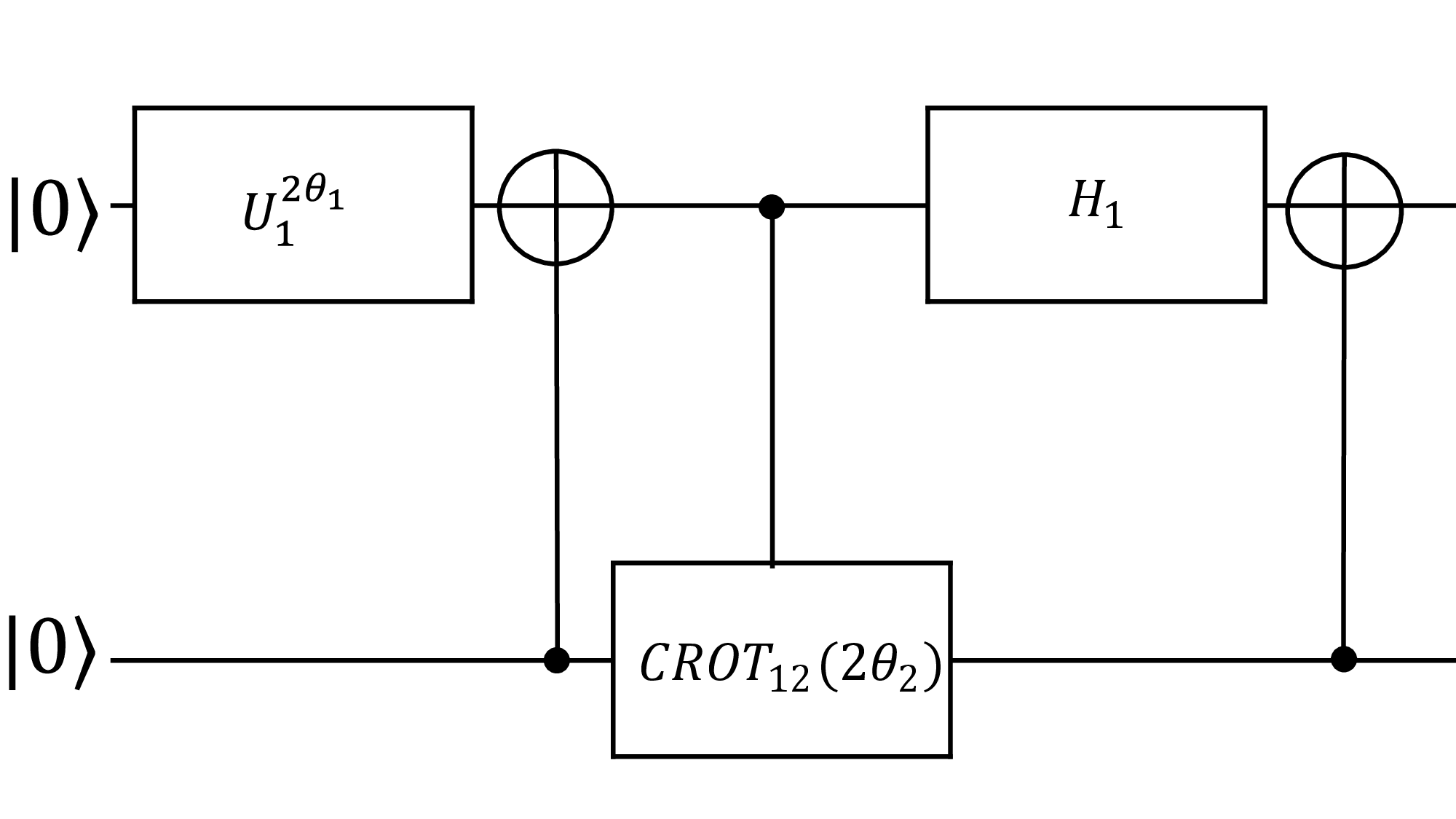}
	\caption{Quantum circuit \cite{shruti_2015} showing the sequence of implementation of single-qubit unitary transformation, two-qubit controlled NOT gate, two-qubit controlled rotation gate and single-qubit Hadamard gate required for the construction of two-qubit state possessed by Alice, given by (\ref{alice_two_qubit_state_nmr_final}).}
	\label{fig_1}
\end{figure}

$\text{CROT}_{12}^{2\theta_2}$ implements a controlled rotation by an arbitrary angle $\theta_2\in [0,\pi/2)$, with the first qubit as control qubit and second qubit as the target qubit. The final state obtained is dependent upon $\theta_1$ and $\theta_2$.\\	
Now, we choose the values of $\theta_1$ and $\theta_2$ in such a way that the coefficients of the like terms of the state given in (\ref{alice_two_qubit_state_nmr}) matches with the coefficients of the output state obtained in (\ref{nmr_output}). Therefore, we have
\begin{align}
	\cos \theta_1+\sin \theta_1\sin \theta_2&=\cos \theta_1-\sin  \theta_1\sin \theta_2=\alpha\\
	\sin\theta_1 \cos \theta_2&=\beta
\end{align}
Here, we choose $\theta_2=0$, which reduces equation (\ref{nmr_output}) to 
\begin{align}
	\frac{1}{\sqrt{2}}\bigg[\cos \theta_1 (\ket{00}+\ket{11})+\sin \theta_1 (\ket{01}-\ket{10})\bigg]
	\label{alice_two_qubit_state_nmr_final}
\end{align}
From (\ref{alice_two_qubit_state_nmr_final}), we obtain the required values of state parameters as 
\begin{align}
	\begin{split}
		\alpha=\cos \theta_1\\
		\beta=\sin \theta_1
	\end{split}
\end{align}
which satisfies the normalization condition $\alpha^2+\beta^2=1$, for $\theta_1\in (0,\pi/2)$.\\
The form of the two-qubit state used in the BB84-like protocol can be generated in an idealized experimental setup as discussed above. The quantum circuit considered here provides a convenient representation for the generation of the desired states. In experimental practicality, factors such as decoherence, gate imperfections, and detector inefficiencies may introduce additional disturbances in the resultant states. Such factors may affect the performance of the protocol. However, the proposed state preparation procedure gives useful insights in generating the desired two-qubit states that are used in the BB84-like QKD protocol.

\subsection{Role of eavesdropper}

If we assume that a potential eavesdropper, say Eve is present in between the communicating parties, say, Alice and Bob then her role is to secretly intercept Alice's state that is being transferred to Bob. Eve's intention is to extract as much information as possible on the key exchanged between Alice and Bob. To execute her plan of stealing the information, she make use of a quantum cloning machine (QCM) to copy the intercepted state. After successfully applying $1\longrightarrow 2$ quantum cloning transformation $U$, she keeps one copy with herself and sends the other copy to Bob, in order to remain hidden in the key distribution process. \\
%Eve possesses all the powers, which are permitted by the laws of quantum mechanics \cite{gisin_1997}.
Let us consider that Eve's cloning transformation $U$ acts on the intercepted state in the following way:	
\begin{align}
	U(\ket{S}_{in}&\bra{S}\otimes \ket{0}_b \bra{0} \otimes \ket{Q}_x \bra{Q})U^\dagger,~~~S\equiv \phi~~\text{or}~~\phi'
\end{align}
where $\ket{Q}_x$ denotes the QCM state vector, and the subscript $"in"$ and $"b"$ denote the input state and the copy mode, respectively. Eve obtains the copied state by applying a partial trace operation over the machine state vector $\ket{Q}_x$ and the input state $\ket{S}_{in}$ and keeps one of the copied state described by the density operator $\rho_E$ or $\rho_E'$ with herself.\\
%The states obtained by Eve can be calculated by taking partial trace over Bob's bits, and are given by $\rho_E=\text{Tr}_B[U\ket{\phi}\bra{\phi}]$ and $\rho_E'=\text{Tr}_B[U\ket{\phi'}\bra{\phi'}]$.\\
Since the sender Alice and the receiver Bob have no idea about the existence of an eavesdropper in between so, Bob, after receiving the state may proceed with the protocol if the key rate $r>0$, or they abort it and start over again if the key rate $r<0$ \cite{wang_2023, sohr_2024}. The key rate $r$ depends upon the \textquotedblleft $Z$ basis\textquotedblright~error rate $\delta_z$ and it may increase due to the noise present in the communication channel or the interference of the eavesdropper who is present in between the communicating parties. Thus, it would be interesting to study the effect of Eve's cloning transformation on $\delta_z$ which may lead to the failure of the protocol and thus may reveal Eve's presence in the key distribution process. Hence, our task is to explore the possible bound of the key rate where the key distribution protocol is successful, even when an eavesdropper is present in the protocol. 

\section{Lower bound for a positive key rate}

In this section, we would like to explore the possible lower bound of the secret key rate when the key distribution protocol is successful, even in the presence of the eavesdropper. Shor and Perskill gave the lower bound on the key rate $r$, which is given by \cite{shor_2000}
\begin{align}
	r\geq 1-h(\delta_x)-h(\delta_z)
	\label{r bound shor}
\end{align}
where $\delta_x$ and $\delta_z$ are the $X$ and $Z$ basis error rates, respectively. Woodhead modified the lower bound (\ref{r bound shor}) on $r$ using the cloning bound, for a variant of BB84 scheme, which is given by \cite{woodhead_2013}
\begin{align}
	r\geq 1-h(u)-h(\delta_z)
	\label{r bound woodhead}
\end{align}
where $u=\frac{1}{2}(1+F(\rho_E,\rho_E'))$ and $h(x)=-x \log_2 (x)-(1-x) \log_2 (1-x)$, is the binary (Shannon) entropy function \cite{cover_1999_book}.  The fidelity $F(\rho_E,\rho_E')$ measures the overlapping between the output states $\rho_E$ and $\rho_E'$ obtained after applying the cloning transformation on the two input states $\ket{\phi}$ and $\ket{\phi'}$, respectively at Eve's site. The fidelity $F(\rho_E,\rho_E')$ can be calculated as $F(\rho_E,\rho_E')=||\sqrt{\rho_E}\sqrt{\rho_E'}||_1$, where $||\cdot||_1$ denotes the trace norm.\\
%Woodhead estimated the lower bound of the secret key rate $r$, and our task is to determine the upper and lower bound of this estimate. The lower bound (\ref{r bound woodhead}) is based on Eve's choice of the cloning machine and the error rate $\delta_z$.
% As discussed in the previous section, Eve uses a quantum cloning transformation, and it may affect the key distribution process by affecting the key rate $r$. Therefore, using the Devetak-Winter bound \cite{devetak_2005}, a lower bound on $r$ can be established in terms of fidelity $F(\rho_E,\rho_E')$, which is a measure to calculate distance between two states, and the error rate $\delta_z$. The fidelity $F(\rho_E,\rho_E')$ depends only upon the parts of Z basis the eavesdropper can access.
It is interesting to note that the right hand side of (\ref{r bound woodhead}) might be positive or negative, based on the values of $h(u)$ and $h(\delta_z)$. A key distribution protocol is considered to be successful if the key rate is positive, i.e., $1-h(u)-h(\delta_z)>0$.	Therefore, for a successful QKD protocol, we aim to find the lower bound for $1-h(u)-h(\delta_z)$ such that the key rate $r$ is always positive.
\begin{theorem}
	The lower bound of the secret key rate $r$, for a successful key distribution scheme can be given by
	\begin{align}
		R_{lb}&> \frac{1}{\log 2} \left[a +\frac{\delta_z^2-1}{2.5} -\delta_z \right]
	\end{align}		
	where $R_{lb}>r$, and the corresponding range for the error rate $\delta_z$ can be given as
	\begin{align}
		0<\delta_z<\frac{2.5- \sqrt{10.25-10a}}{2}
	\end{align}
\end{theorem}
\begin{proof} For the sake of simplicity, let us assume that $1-h(u)-h(\delta_z)\equiv R$. We find the lower bound of $R$ denoted by $R_{lb}$ such that $r$ is always positive, i.e., $0<R_{lb}<R<r$. Now, expressing $R$ in terms of fidelity $F(\rho_E,\rho_E')$ and $\delta_z$, we get
	\begin{widetext}
		\begin{small}
			\begin{align}
				\begin{split}
					R=&\frac{1}{\log 2} \left[\log 2+\left(\frac{1+F(\rho_E,\rho_E')}{2}\right)\log \left(\frac{1+F(\rho_E,\rho_E')}{2}\right)+ \left(\frac{1-F(\rho_E,\rho_E')}{2}\right) \log \left( \frac{1-F(\rho_E,\rho_E')}{2}\right)+\delta_z \log \delta_z +(1-\delta_z) \log (1-\delta_z)\right]
					%1+\left(\frac{1+F(\rho_E,\rho_E')}{2}\right)\log_2 \left(\frac{1+F(\rho_E,\rho_E')}{2}\right)+ \left(\frac{1-F(\rho_E,\rho_E')}{2}\right) \log_2 \left( \frac{1-F(\rho_E,\rho_E')}{2}\right)+\delta_z \log_2 \delta_z +(1-\delta_z) \log_2 (1-\delta_z)\\
				\end{split}	
				\label{R_log2}
			\end{align}
		\end{small}
	\end{widetext}
	%\subsection{Lower bound on $R$}
	In order to find the lower bound on $R$ such that $0<R_{lb}<R$, we rewrite (\ref{R_log2}) as 
	\begin{align}
		R=&\frac{1}{\log 2} [a +\delta_z \log \delta_z +(1-\delta_z) \log (1-\delta_z)]
		\label{r_lb_step1}
		%\end{widetext}
	\end{align}
	where $a=\log 2 + \left(\frac{1+F(\rho_E,\rho_E')}{2}\right)\log \left(\frac{1+F(\rho_E,\rho_E')}{2}\right)+ \left(\frac{1-F(\rho_E,\rho_E')}{2}\right) \log \left( \frac{1-F(\rho_E,\rho_E')}{2}\right)$.\\
	Let us consider the inequality 
	\begin{align}
		\dfrac{x}{1+x}\leq \log (1+x), ~x>-1  
		\label{ineq_x/1+x}
	\end{align}
	We will use this inequality when $-1<x<0$ and by replacing $x$ with $-\delta_z$, the inequality (\ref{ineq_x/1+x}) reduces to
	\begin{align}
		-\delta_z\leq (1-\delta_z) \log (1-\delta_z), ~ 0<\delta_z<1
		\label{ineq_dz_logdz_lb}
	\end{align}
	Using the inequality (\ref{ineq_dz_logdz_lb}) in the expression (\ref{r_lb_step1}), we get
	\begin{align}
		\begin{split}
			R&\geq \frac{1}{\log 2} \left[a +\delta_z \log \delta_z -\delta_z \right] \equiv R_{lb}
			\label{lb_rlb_a dz}
		\end{split}	
	\end{align}		
	The above inequality (\ref{lb_rlb_a dz}) is very complicated to solve analytically. Therefore, for simplicity of the calculations, we express the term $\log \delta_z$ as an inequality involving polynomial terms of $\delta_z$. Therefore, applying the inequality $\log \delta_z > \dfrac{\delta_z^2-1}{2.5\delta_z}$ for $\delta_z\in (0,0.305)$, the expression of $R_{lb}$ given in (\ref{lb_rlb_a dz}) further reduces to 
	\begin{align}
		R_{lb}&> \frac{1}{\log 2} \left[a +\frac{\delta_z^2-1}{2.5} -\delta_z \right]
		\label{lb_Rlb}
	\end{align}		
	It can be noted here that for different choices of the constant other than $2.5$, we may have different intervals of $\delta_z$. There may exist values of the constant other than $2.5$ that may give a range of the error rate for a successful key distribution protocol. However, in a QKD protocol, smaller error rates are preferable since they correspond to a lower disturbance in the communication channel and result in successful generation of the secret key.\\ %The choice of $2.5$ provides a mathematically valid approximation for a successful key distribution scheme and results in reasonable low error rates.
	For $R_{lb}>0$, we have to find the values of fidelity $F(\rho_E,\rho_E')$ and $\delta_z$ for which the right hand side of (\ref{lb_Rlb}) is positive, i.e., $a +\dfrac{\delta_z^2-1}{2.5} -\delta_z>0$ holds.\\
	Here, we have considered $F(\rho_E,\rho_E')$ and $\delta_z$ as variables, rather than choosing a specific value for them. As the input states $\rho_E$ and $\rho_E'$ changes, $F(\rho_E,\rho_E')$ also changes. Therefore, by using the concept of symbolic computation \cite{gao_2024,gao_2025}, we find the possible ranges of $F(\rho_E,\rho_E')$ and $\delta_z$ such that $R_{lb}>0$ is satisfied. Simplifying this condition, we get a quadratic equation in $\delta_z$, which is given by
	\begin{align}
		\delta_z^2-2.5 \delta_z +2.5a -1>0
		\label{ineq_lb_dz a}
	\end{align}
	Solving the inequality (\ref{ineq_lb_dz a}), we get 
	\begin{align}
		0<\delta_z<\frac{2.5- \sqrt{10.25-10a}}{2}
		\label{delta_z_bound_R_lb}
	\end{align}
	and
	\begin{align}
		\frac{2.5+ \sqrt{10.25-10a}}{2}<\delta_z
		\label{eliminate_2}
	\end{align}
	Since, in general, $\delta_z\in (0,1)$, so we eliminate (\ref{eliminate_2}). Therefore, the inequality (\ref{delta_z_bound_R_lb}) holds for $F(\rho_E,\rho_E')\in (0.8281,0.9922)$.
\end{proof}
Substitute $a$ in terms of $F(\rho_E,\rho_E')$ in the inequality (\ref{delta_z_bound_R_lb}) and solving, we find that for a given $F(\rho_E,\rho_E')$, $\delta_z\in (0,\delta_{z_1})$, where $\delta_{z_1}\in (0, 0.305)$. Therefore, the inequality $0<R_{lb}<R$ holds when $\delta_z\in (0,\delta_{z_1})$, where $\delta_{z_1}\in (0, 0.305)$. Thus, in the given range of $F(\rho_E,\rho_E')$ and $\delta_z$, we find that the key rate $r$ is always positive and hence it results into the success of the key distribution protocol. 

\section{Analysis of various symmetric QCMs}
In this section, we would like to study the effect of Eve's cloning transformation on the secret key rate $r$. The success and failure of the protocol depend on its estimate $R$ given in (\ref{R_log2}), which can be positive or negative, respectively. The estimated value of $R$ depends on the type of cloning transformation used by the eavesdropper. Eve clone the input states $\ket{\phi}$ and $\ket{\phi'}$ and obtains the output states $\rho_E$ and $\rho_E'$, respectively using various symmetric QCMs. The overlapping between these output states $F(\rho_E,\rho_E')$ depends on the input state parameters $\alpha$ and the cloning machine parameters. We, therefore, study various types of symmetric QCMs used by the eavesdropper and derive the relationship between $F(\rho_E,\rho_E')$, the input state parameters, and the cloning machine parameters. For the given values of the input state parameters and the cloning machine parameters, we find the values of $F(\rho_E,\rho_E')$ and analyze the value of $R$ for the corresponding range of $\delta_z$ given in (\ref{delta_z_bound_R_lb}). If $R_{lb}>0$, then the secret key can be established successfully between two distant partners Alice and Bob. 
%    Based on the concept of symmetric QCMs, there are input-state-dependent and input-state-independent QCMs. \\
% We first consider the general case of input state-dependent Wootters-Zurek QCM. We shall compare this cloning transformation with the proposed input state independent QCM. \\
% Further, we would like to study about the role of efficiency of the applied QCM for a successful key distribution protocol.

\subsection{Wootter-Zurek QCM}
\begin{figure}[H]
	\centering
	\includegraphics[width=1\linewidth]{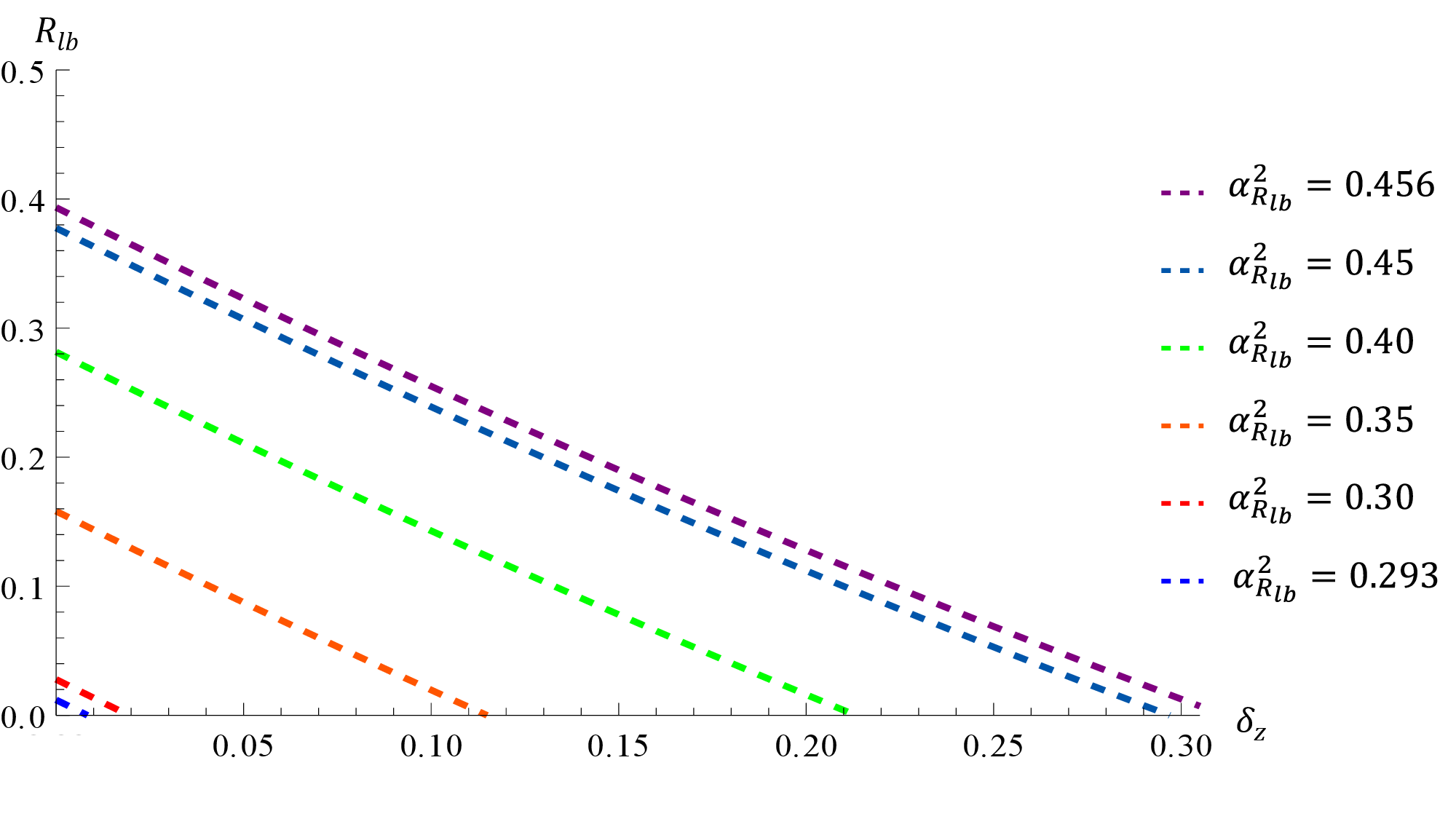}
	\caption{The lower bound of $R$ for a successful QKD protocol ($r>0$) when the eavesdropper uses WZ QCM to extract information from the intercepted state is shown. The $x-$ axis represents the range of $\delta_z$, for a given value $F_{WZ}(\rho_E,\rho_E')$, which depends only on Alice's secret state parameter $\alpha^2\in(0.293,0.456)$. The $y-$ axis represents the value of lower bound $R_{lb}$ such that $0<R_{lb}<R$.}
	\label{fig_2}
\end{figure}

In 1982, Wootters and Zurek \cite{wooters_1982} analyzed the copying process of an arbitrary input state $\ket{\phi}_a$ which is a linear combination of basis states vectors $\ket{0}_a$ and $\ket{1}_a$, the subscript $a$ denotes the original system. The cloning transformation is defined on the basis vectors $\ket{0}$ and $\ket{1}$, and is given by
\begin{align}
	\begin{split}
		\ket{0}_a\ket{0}_b\ket{Q}_x \longrightarrow \ket{0}_a\ket{0}_b\ket{Q_0}_x\\
		\ket{1}_a\ket{0}_b\ket{Q}_x \longrightarrow \ket{1}_a\ket{1}_b\ket{Q_1}_x
		\label{wz_qcm}
	\end{split}
\end{align}
where $\ket{Q}_x$ denotes the QCM state vector and, the subscript $b$ denotes the copy mode of the input state. The copying state vectors satisfies the conditions 
\begin{align}
	\braket{Q_0}{Q_1}=0, \braket{Q_i}{Q_i}=1, i=0,1
\end{align} 
The WZ copying machine is input state-dependent, i.e., the quality of the copies at the output of the quantum cloner depends on the input state.\\
Let us assume that Eve uses WZ QCM as described by (\ref{wz_qcm}) to copy the intercepted state that Alice wishes to send to Bob. Therefore, the cloning transformation can be read as
\begin{align}
	\begin{split}
		\ket{0}_A\ket{0}_E\ket{Q}_{M} \longrightarrow \ket{0}_B\ket{0}_E\ket{Q_0}_{M}\\
		\ket{1}_A\ket{0}_E\ket{Q}_{M} \longrightarrow \ket{1}_B\ket{1}_E\ket{Q_1}_{M}
		\label{wz_qcm_eve}
	\end{split}
\end{align}
where the subscript $A$ refers to Alice's qubit that Eve intercepts, the subscript $B$ refers to the qubit that Eve sends to Bob after copying the intercepted state in order to remain hidden in the key distribution process, Eve's copy state is expressed in the basis state $\ket{i}_E, i=0,1$, and $\ket{Q_i}_{M}$ refers to Eve's cloning machine state vectors at the output.\\
Let us assume that Alice sends the state $\ket{\phi}=\alpha \ket{0}+\beta \ket{1}$ to Bob. Eve intercepts the state in the midway and applies WZ QCM given by (\ref{wz_qcm_eve}) and obtains the state at the output as $\rho_{{BEM}}$. The two-qubit copy state described by the reduced density operator $\rho_{BE}$ after applying cloning operation is given by 
\begin{align}
	\rho_{BE}=\text{Tr}_{M}[\rho_{BEM}]=\alpha^2 \ket{00}_{BE}\bra{00}+\beta^2\ket{11}_{BE}\bra{11}
\end{align}	
The single-qubit state $\rho_E$, which is in possession of Eve is given by
\begin{align}
	\rho_E=\text{Tr}_{B}[\rho_{BE}]=\alpha^2 \ket{0}\bra{0}+\beta^2 \ket{1}\bra{1}
	%		\begin{bmatrix}
		%			\alpha^2 &  0 \\
		%			0 & \beta^2
		%		\end{bmatrix}
	\label{wz_rhoe}
\end{align}
Further, if Alice sends the qubit $\ket{\phi'}=\alpha \ket{1}-\beta \ket{0}$, then Eve performs the same cloning operation on the intercepted qubit, which results in the two-qubit state after eliminating the machine state vector which is given by 
\begin{align}
	\rho_{BE'}=\text{Tr}_{M}[\rho_{BE'M}]=\beta^2 \ket{00}_{BE'}\bra{00}+\alpha^2\ket{11}_{BE'}\bra{11}
\end{align}
Eve keeps the single-qubit state with herself and it is described by the density operator $\rho_{E}'$, which is given by
\begin{align}
	\rho_E'=\beta^2\ket{0}\bra{0}+\alpha^2\ket{1}\bra{1}
	%		\begin{bmatrix}
		%			\beta^2 &  0 \\
		%			0 & \alpha^2 
		%		\end{bmatrix}
	\label{wz_rhoe'}
\end{align}
The overlapping between two density matrices $\rho_E$ and $\rho_E'$ can be described as \cite{muller_2023}
\begin{align}
	F(\rho_E,\rho_E')=\bigg(\sum_{i} \sqrt{\lambda_i(\rho_E,\rho_E')}\bigg)^2
\end{align}
where $\lambda_i$ is the $i^{th}$ eigenvalue of the product of density matrices $\rho_E$ and $\rho_E'$. \\
Therefore, the fidelity between Eve's output states $\rho_E$ and $\rho_E'$ when she uses WZ QCM is given by (See Appendix B.1 for calculations (\ref{wz_fid_a1}))
\begin{align}
	F_{WZ}(\rho_E,\rho_E')=4\alpha^2 (1-\alpha^2)
	\label{wz_fid}
\end{align}
It can be observed that $F_{WZ}(\rho_E,\rho_E')$ only depends upon Alice's secret state parameter $\alpha^2$. It is interesting to note here that $F_{WZ}(\rho_E,\rho_E')$ displays a symmetric behavior when $\alpha^2$ is lying in the interval $(0,0.5)$ and $(0.5,1)$. Therefore, we will only consider the values of $\alpha^2 \in (0,0.5)$. Further, we can find the values of $\alpha^2$, for which the lower bound $R_{lb}>0$ is satisfied i.e., the key distribution protocol is successfully completed with $R>0$. To achieve this, we have analyzed the values of $\alpha^2$ and found that when $\alpha^2\in (0.293,0.456)$, the fidelity  $F_{WZ}(\rho_E,\rho_E')\in (0.8281,0.9922)$. With the help of the range of $F_{WZ}(\rho_E,\rho_E')$, we find the corresponding range of $\delta_z$ for which $R_{lb}>0$. This is depicted in fig \ref{fig_2}.	

It can be observed from fig \ref{fig_2} that as Alice increases the value of her secret state parameter $\alpha^2\in (0.293,0.456)$, the overlapping between Eve's output states $F(\rho_E,\rho_E')$ also increases when Eve uses WZ QCM. For an increasing $F(\rho_E,\rho_E')$, the range of $\delta_z$ also increases significantly. This is valid because increasing fidelity implies better cloning, and therefore increase in error rate. The estimate of the lower bound of the key rate $r$, given by $R_{lb}$ varies accordingly.\\
Since, the protocol is successful in generating the secret key for $R_{lb}>0$, so Alice is restricted to choosing the state parameter $\alpha^2$ in the interval $(0.293,0.456)$. Otherwise, there might arise a case when Alice chooses $\alpha^2\in (0,0.293)\cup (0.456,0.5)$ then $R_{lb}$ might be negative, and thus resulting in the failure of the key distribution protocol. Therefore, the question arises that whether there exist some other cloning machine that may be used by the eavesdropper for which Alice can choose the secret parameter $\alpha^2$ in the interval $(0,1)$ with $R_{lb}>0$ and even then Eve manages to remain hidden in the key distribution protocol?

%	Since, Eve's fidelity $F_{WZ}(\rho_E,\rho_E')$ depends only upon $\alpha^2$, so Alice might be able to choose $\alpha^2$ such that $F(\rho_E,\rho_E')\rightarrow 0$. Naturally, the question arises whether we can have any value of $\alpha^2$ for which $F(\rho_E,\rho_E')\rightarrow 0$. This problem can not be addressed by WZ cloning machine, so we have to look for some other cloning machine which can provide an answer to this problem.

\subsection{Modified Buzek-Hillery symmetric QCM}
In this subsection, we modified the QCM prescribed by Buzek and Hillery \cite{buzek_1996} and checked whether it fulfilled the asked question in the previous subsection. We modify the BH QCM such that it maybe used by Eve to clone Alice's intercepted qubit such that $R_{lb}>0$ and she remains hidden in the key distribution protocol.\\
In 1996, Buzek and Hillery introduced the input state independent QCM, which has the property that the quality of the output state was independent of the input state. \\
The BH quantum cloning transformation is given by
%\begin{widetext}
\begin{equation}
	\begin{split}
		\ket{0}_A\ket{0}_E\ket{Q}_x \longrightarrow& \ket{0}_B\ket{0}_E\ket{Q_0}_x\\ &+[\ket{0}_B\ket{1}_E+\ket{1}_B\ket{0}_E]\ket{Y_0}_x\\
		\ket{1}_A\ket{0}_E\ket{Q}_x \longrightarrow& \ket{1}_B\ket{1}_E\ket{Q_1}_x\\
		&+[\ket{0}_B\ket{1}_E+\ket{1}_B\ket{0}_E]\ket{Y_1}_x
	\end{split}	
	\label{bh_qcm}
\end{equation}
%\end{widetext}
The unitarity of the transformation gives
\begin{align}
	\braket{Q_i}{Q_i}+2\braket{Y_i}{Y_i}=1, i=0,1\\
	\braket{Y_i}{Y_j}=\braket{Q_i}{Q_j}=0, i\neq j, i,j=0,1\\
	%\braket{Y_0}{Y_1}=\braket{Y_1}{Y_0}=0
	\braket{Q_i}{Y_i}=\braket{Y_i}{Q_i}=0, i=0,1
	%\braket{Q_1}{Y_1}=\braket{Q_0}{Y_0}=\braket{Y_0}{Q_0}=\braket{Y_1}{Q_1}=0\\
	%\braket{Q_0}{Q_1}=0
\end{align}
In addition to the above conditions, the following also holds
\begin{align}
	\braket{Y_0}{Y_0}=\braket{Y_1}{Y_1}=\xi, 0\leq \xi \leq 1/2\\
	\braket{Y_i}{Q_j}=\braket{Q_i}{Y_j}=\eta/2, i\neq j, i,j=0,1
\end{align}
where $0\leq \eta \leq 1\sqrt{2}$. Here $\xi$ and $\eta$ denotes the cloning transformation parameters. It can be noted here that if we choose $\xi=0$, then it implies $\eta=0$ also, and BH cloning transformation reduces to WZ QCM. So, here we modify the BH QCM by choosing $\eta=0$ and $\xi\neq 0$. In this case, BH QCM will become a state dependent QCM.\\
Let us assume that Eve uses the modified QCM for cloning the intercepted state. Suppose that Alice sends the state $\ket{\phi}=\alpha \ket{0}+\beta \ket{1}$ to Bob, which is intercepted by Eve in the mid way. If she applies the modified BH quantum cloning transformation described in (\ref{bh_qcm}) on the intercepted state then the combined state becomes $\rho_{BEM}$. The resultant single-qubit state $\rho_E$ with Eve, after tracing out the copy qubit and the cloning machine state vector, is given by
\begin{align}
	\rho_E= \alpha^2 +\xi (\beta^2-\alpha^2) \ket{0}\bra{0}+ \beta^2+\xi (\alpha^2-\beta^2) \ket{1}\bra{1}
	%		\begin{bmatrix}
		%			\alpha^2 +\xi (\beta^2-\alpha^2)&  0 \\
		%			0 & \beta^2+\xi (\alpha^2-\beta^2) 
		%		\end{bmatrix}
	\label{our_rhoe}
\end{align}
Similarly, if Alice sends $\ket{\phi'}=\alpha \ket{1}-\beta \ket{0}$ to Bob, then Eve may intercept it and applies the modified BH quantum cloning transformation (\ref{bh_qcm}) on it. In this case, the single-qubit state with Eve is given by
\begin{align}
	\rho_E'=\beta^2+\xi (\alpha^2-\beta^2) \ket{0}\bra{0}+\alpha^2 +\xi (\beta^2-\alpha^2) \ket{1}\bra{1}
	%		\begin{bmatrix}
		%			\beta^2+\xi (\alpha^2-\beta^2) &  0 \\
		%			0 & \alpha^2 +\xi (\beta^2-\alpha^2)
		%		\end{bmatrix}
	\label{our_rhoe'}
\end{align}
The overlap between the two mixed states described by the density operator $\rho_E$ and $\rho_E'$, when Eve uses the modified BH QCM can be given by (Refer Appendix B.2 for calculations (\ref{fid_our_qcm1_a1}))
\begin{align}
	\begin{split}
		F_{{BH}}(\rho_E,\rho_E')=&4[\alpha^2 (1-\alpha^2)(1-2\xi)^2+\xi(1-\xi)]
		\label{fid_our_qcm1}
	\end{split}
\end{align}
It can be clearly seen here that $F_{{BH}}(\rho_E,\rho_E')$ not only depends upon Alice's secret state parameter $\alpha^2$, but also on the cloning machine parameter $\xi$. \\
%Here, even if Alice cleverly chooses a value of $\alpha^2\in(0,1)$, $F_{BH}(\rho_E,\rho_E')$ will vary depending on $\xi\in (0,0.5]$. \\
The expression of $F_{BH}(\rho_E,\rho_E')$ obtained in (\ref{fid_our_qcm1}) lies in the interval $(0,1]$, for $\alpha^2\in (0,1)$ and $\xi \in (0,0.5]$. In this case also, we find that $F_{BH}(\rho_E,\rho_E')$ is a symmetric function with respect to the parameter $\alpha^2$. Alice choose the value of $\alpha^2$ to prepare her secret state without having any idea about the cloning machine state parameter $\xi$, used by Eve in the construction of modified BH QCM to extract information from the intercepted state. We have shown that for a given value of $\xi\in(0,0.455)$, there exists certain values of $\alpha^2 \in(0,0.455)$ for which Eve manages to remain hidden in the key distribution process, and the key is distributed successfully between the sender Alice and receiver Bob. 
\begin{table}[h]
	\scriptsize
	\resizebox{\columnwidth}{!}{%
		\begin{tabular}{c|c|c|c|c|c}
			\cline{2-5}
			& \textbf{$\xi$} & \textbf{$\alpha^2$} & \textbf{$F_{BH}(\rho_E,\rho_E')$} & \textbf{$\delta_z$} &  \\ \cline{2-5}
			& \multirow{5}{*}{0.1} & 0.241 & 0.8283 & (0,0.0003) &  \\ \cline{3-5}
			&  & 0.30 & 0.8976 & (0,0.0947) &  \\ \cline{3-5}
			&  & 0.35 & 0.9424 & (0,0.1748) &  \\ \cline{3-5}
			&  & 0.40 & 0.9744 & (0,0.2495) &  \\ \cline{3-5}
			&  & 0.445 & 0.9936 & (0,0.3049) &  \\ \cline{2-5}  
			& \multirow{7}{*}{0.2} & 0.155 & 0.8286 & (0,0.0006) &  \\ \cline{3-5} 
			&  & 0.20 & 0.8704 & (0,0.0543) &  \\ \cline{3-5}
			&  & 0.25 & 0.9100 & (0,0.1149) &  \\ \cline{3-5}
			&  & 0.30 & 0.9424 & (0,0.1748) &  \\ \cline{3-5}
			&  & 0.35 & 0.9676 & (0,0.2318) &  \\ \cline{3-5}
			&  & 0.40 & 0.9856 & (0,0.2823) &  \\ \cline{3-5}
			&  & 0.426 & 0.9921 & (0,0.3044) &  \\ \cline{2-5}
			& \multirow{9}{*}{0.3} & 0.001 & 0.8406 & (0,0.0153) &  \\ \cline{3-5}
			&  & 0.05 & 0.8704 & (0,0.0544) &  \\ \cline{3-5}
			&  & 0.10 & 0.8976 & (0,0.0947) &  \\ \cline{3-5}
			&  & 0.15 & 0.9216 & (0,0.1350) &  \\ \cline{3-5}
			&  & 0.20 & 0.9424 & (0,0.1748) &  \\ \cline{3-5}
			&  & 0.25 & 0.9600 & (0,0.2133) &  \\ \cline{3-5}
			&  & 0.30 & 0.9744 & (0,0.2495) &  \\ \cline{3-5}
			&  & 0.35 & 0.9856 & (0,0.2823) &  \\ \cline{3-5}
			&  & 0.39 & 0.9923 & (0,0.3049) &  \\ \cline{2-5}
			& \multirow{7}{*}{0.4} & 0.001 & 0.9602 & (0,0.2137) &  \\ \cline{3-5}
			&  & 0.05 & 0.9676 & (0,0.2318) &  \\ \cline{3-5}
			&  & 0.10 & 0.9744 & (0,0.2496) &  \\ \cline{3-5}
			&  & 0.15 & 0.9804 & (0,0.2665) &  \\ \cline{3-5}
			&  & 0.20 & 0.9856 & (0,0.2823) &  \\ \cline{3-5}
			&  & 0.25 & 0.9900 & (0,0.2969) &  \\ \cline{3-5}
			&  & 0.28 & 0.9923 & (0,0.3049) &  \\ \cline{2-5}
			& \multirow{2}{*}{0.455} & 0.001 & 0.9919 & (0,0.3038) &  \\ \cline{3-5}
			&  & 0.011 & 0.9923 & (0,0.3049) &  \\ \cline{2-5}
		\end{tabular}%
	}
	\caption{For a given value of $\xi\in (0,0.455)$, there exists values of Alice's secret state parameter $\alpha^2$ such that the lower bound required for a successful QKD protocol, i.e., $R_{lb}>0$ is achieved, even when Eve uses the BH QCM.}
	\label{table_mbh_qcm}
\end{table}
\\The results in the given table \ref{table_mbh_qcm} can be validated by considering the fact that as the fidelity for Eve increases, the interval of the error rate also increases due to Eve's intervention. We have shown that, it is possible to form the secret key between the sender and receiver, in spite of the increasing interval of the error rate.
%\\ increases and thus, error rate also in Alice and Bob's key distribution increases, but they still be able to distill a key.\\
%Since the input mode is not equal to the output mode, we are now eager to find the distance between the input mode and the output mode of a QCM. Now we are in a position to ask the following questions: (i) How efficient can a QCM be in cloning an arbitrary input state and, (ii) Does there exists a relationship between the efficiency of the QCM and the fidelity $F(\rho_E,\rho_E')$? 

\section{Efficiency of the QCM}

Alice prepares her secret states $\ket{\phi}$ or $\ket{\phi'}$ and then sends it to Bob. Eve intercepts Alice's state in the midway, before reaching to Bob. She then applies a quantum cloning transformation to extract the information from the intercepted state. After applying the quantum cloning transformation on the intercepted state $\ket{\phi}$ or $\ket{\phi'}$, the possible output is represented by the density operator $\rho_E$ or $\rho_E'$, respectively. \\
In this section, we will study the efficiency of the QCM by calculating the Hilbert-Schmidt (HS) distance between the input (ideal) state, described by the density operator $\rho_{id}$ or $\rho_{id}'$ and the output state of the cloning machine, described by the density operator $\rho_E$ or $\rho_E'$. We have shown that the HS distance between $\rho_{id}$ ($\rho_{id}'$) and $\rho_E$ ($\rho_E'$) is upper bounded by the trace distance between $\rho_E$ and $\rho_E'$, by using the relationship between trace distance and the fidelity. Finally, we have analyzed the inequality between the HS distance and the trace distance to find the effect of the efficiency of the QCM on the successful key generation, i.e., $R_{lb}>0$. \\
The distance between the density matrices $\rho^{in}$ and $\rho^{out}$ can be calculated by trace distance $D(\rho^{in},\rho^{out})$, defined as 
\begin{align}
	D(\rho^{in},\rho^{out})=\frac{1}{2} \text{Tr}|\rho^{in}-\rho^{out}|=\frac{1}{2}||\rho^{in}-\rho^{out}||_1
\end{align}
where $||M||_1=\text{Tr}\sqrt{M^\dagger M}$ denotes the trace norm for an arbitrary operator $M$. In higher dimensional systems, trace distance might require diagonalization of density matrices, and computing the diagonalized matrix can be exponentially difficult, making trace distance a less preferred choice \cite{dononov_2000}. To overcome this difficulty, an alternative measure is the Hilbert-Schmidt (HS) distance \cite{witte_1999, ozawa_2000, jain_2025}, which can be defined as
\begin{align}
	D_{HS}(\rho,\sigma)=\text{Tr}[(\rho-\sigma)^2]=||\rho-\sigma||_2^2
\end{align}
where $\rho$ and $\sigma$ are two arbitrary quantum states. $||M||_2=\sqrt{\text{Tr}(M^\dagger M)}$ denotes the Frobenius norm which is easier to calculate both numerically and analytically. Furthermore, employing HS distance can be cost efficient as it can be efficiently computed on a quantum computer, thereby reducing the cost of the experiment \cite{cincio_2018}. It provides a direct interpretation as an information distance between quantum states \cite{li_2003}. The efficiency of a QCM can be analyzed using the HS distance.

\subsection{Relation between the efficiency of the QCM and the secret key rate for a successful QKD protocol}
To establish the relation between the efficiency of the QCM and the generation of the secret key, we start with the relationship between trace distance $D(\rho_E,\rho_E')$ and fidelity $F(\rho_E,\rho_E')$, which is given by \cite{neilsen_chuang}
\begin{align}
	D(\rho_E,\rho_E')^2\leq 1-F(\rho_E,\rho_E')^2
	\label{complementary_cond}
\end{align}
\begin{theorem}
	The relationship between the HS distance of $\rho_E$ ($\rho_E'$) and $\rho_{id}$ ($\rho_{id}'$) with the trace distance between $\rho_E$ and $\rho_E'$ can be given by
	\begin{align}
		D_{HS}(\rho_E,\rho_{id})\leq 2 D(\rho_E,\rho_E')^2
	\end{align}
	and 
	\begin{align}
		D_{HS}(\rho_E',\rho_{id}')\leq 2 D(\rho_E,\rho_E')^2
	\end{align}
\end{theorem}
\begin{proof}
	In 2019, Coles et. al. proposed a strong bound between the trace distance $D(\rho,\sigma)$ and the HS distance $D_{HS}(\rho,\sigma)$ between the two density matrices $\rho$ and $\sigma$ as \cite{coles_2019}
	\begin{align}
		\frac{1}{2} D_{HS}(\rho,\sigma)\leq D(\rho,\sigma)^2
		\label{dhs_d_rel}
	\end{align} 
	Taking $\rho \equiv \rho_E$ and $\sigma \equiv \rho_E'$, the inequality (\ref{dhs_d_rel}) reduces to
	\begin{align}
		D_{HS}(\rho_E,\rho_E')\leq 2 D(\rho_E,\rho_E')^2
		\label{dhs_d_rel_qcm}
	\end{align} 
	Using the triangular inequality of the distance measure, we get 
	%Since $D_{HS}$ is a distance measure, it satisfies the triangular inequality between $\rho_{id}$, $\rho_E$ and $\rho_E'$, given by
	\begin{align}
		&D_{HS}(\rho_E,\rho_E')\leq D_{HS}(\rho_E,\rho_{id})+D_{HS}(\rho_E',\rho_{id})\notag \\
		\implies &D_{HS}(\rho_E,\rho_E')-D_{HS}(\rho_E,\rho_{id})\leq D_{HS}(\rho_E',\rho_{id})
	\end{align}
	Since, $D_{HS}(\rho_E',\rho_{id})\geq 0$, so we impose the condition 
	\begin{align}
		D_{HS}(\rho_E,\rho_{id})\leq  D_{HS}(\rho_E,\rho_E')
		\label{ineq_dhs}
	\end{align}
	Using (\ref{dhs_d_rel_qcm}) and (\ref{ineq_dhs}), we have
	\begin{align}
		D_{HS}(\rho_E,\rho_{id})\leq 2 D(\rho_E,\rho_E')^2
		\label{dhs_d_rel_fin}
	\end{align} 
	Similarly, we get
	\begin{align}
		D_{HS}(\rho_E',\rho_{id}')\leq 2 D(\rho_E,\rho_E')^2
		\label{dhs_d1_rel_fin}
	\end{align} 
	Equations (\ref{dhs_d_rel_fin}) and (\ref{dhs_d1_rel_fin}) are the required relationships between the HS distance between $\rho_E$ ($\rho_E'$) and $\rho_{id}$ ($\rho_{id}'$) and the trace distance between $\rho_E$ and $\rho_E'$. 
\end{proof}
We find from equation (\ref{delta_z_bound_R_lb}), that if $F(\rho_E,\rho_E')\in (0.8281,0.9922)$, then $0<R_{lb}<R$, i.e., the secret key can be formed between Alice and Bob. Then for a given value of $F(\rho_E,\rho_E')\in (0.8281,0.9922)$, we have $\delta_z\in(0,\delta_{z_1})$, where $\delta_{z_1}\in (0,0.305)$. Thus, for a given value of $F(\rho_E,\rho_E')\in (0.8281,0.9922)$, a range of $D(\rho_E,\rho_E')\in (0,0.56058)$ can be obtained by using the inequality given in (\ref{complementary_cond}). Hence, for a given value of $D(\rho_E,\rho_E')$, an upper bound of $D_{HS}(.,.)$ can be obtained. This implies that, for a given fidelity $F(\rho_E,\rho_E')$, we can estimate an upper bound on the efficiency of the QCM. This further suggest us that there is an upper bound of the efficiency of the QCM, beyond which QKD protocol may be aborted.  

%Thus, we have shown that there exist QCMs that may be used by the eavesdropper to extract information, which is required by the communicating parties to generate the secret key, without revealing her presence in the QKD protocol.

%	can use a QCM depending upon the upper bound of the efficiency of QCM required for $R_{lb}>0$.
%	Using this, let us assume that $2D(\rho_E,\rho_E')^2=D_1(\rho_E,\rho_E')$, where $D_1(\rho_E,\rho_E')\in (0,0.6285)$. Now, for a given $D_1(\rho_E,\rho_E')$, relation (\ref{dhs_d_rel_fin}) reduces to the
%	\begin{align}
	%		D_{HS}(\rho_E,\rho_{id})\leq D_2(\rho_E,\rho_E')
	%	\end{align}
%	where, $D_2(\rho_E,\rho_E')\in (0,0.6285)$.\\
%	
%	Similar results can be obtained for $D_{HS}(\rho_{id'},\rho_E')$.

\subsection{Analysis of the efficiency of WZ QCM and modified BH QCM and its effect on successful key generation}
In the previous section, we have established the relationship between the HS distance $D_{HS}(.,.)$ and the trace distance $D(.,.)$, which is given by the inequality (\ref{dhs_d_rel_fin}-\ref{dhs_d1_rel_fin}). In this section, we are interested in analyzing the efficiency of two particular QCMs, namely, WZ QCM and modified BH QCM, when the QKD protocol is successful.
%the fidelity $F(\rho_E,\rho_E')$ and  via the trace distance between $\rho_E$ and $\rho_E'$. We are interested in finding out the effect of the fidelity and the efficiency of the cloning machine that maybe used by the Eavesdropper to extract information from the intercepted state in a successful QKD scheme. 
\subsubsection{Efficiency of WZ QCM in successful QKD protocol}
The efficiency of WZ QCM for $R_{lb}>0$ can be analyzed by the table (\ref{table_wz_qcm}) given below.
\begin{table}[h]
	\resizebox{\columnwidth}{!}{%
		\begin{tabular}{|c|c|c|c|}
			\hline
			\textbf{$\alpha^2$} & \textbf{$F_{WZ}(\rho_E,\rho_E')$} & \textbf{$D(\rho_E,\rho_E')^2$} & \textbf{$D_{HS}^{WZ}(\rho_E,\rho_{id})$} \\ \hline
			0.293 & 0.8286 & [0,0.3134) & [0,0.6268) \\ \hline
			0.30 & 0.84 & [0,0.2944) & [0,0.5888) \\ \hline
			0.35 & 0.91 & [0,0.1719) & [0,0.3438) \\ \hline
			0.40 & 0.96 & [0,0.0784) & [0,0.1568) \\ \hline
			0.45 & 0.99 & [0,0.0199) & [0,0.0398) \\ \hline
			0.456 & 0.9922 & [0,0.0154) & [0,0.0311) \\ \hline
		\end{tabular}%
	}
	\caption{Analysis of WZ QCM in Alice's secret state parameter $\alpha^2$ and the HS distance $D_{HS}(\rho_{id},\rho_E)$ between the input and the output mode.}
	\label{table_wz_qcm}
\end{table}

It can be clearly observed from table \ref{table_wz_qcm} that as the fidelity $F(\rho_E,\rho_E')$ increases from 0.8286 to 0.9922, the upper bound of HS distance between $\rho_{id}~(\rho_{id}')$ and $\rho_E~(\rho_E')$ reduces rapidly. This implies that if the eavesdropper uses the WZ QCM to clone the input state $\ket{\phi}$ or $\ket{\phi'}$, then we find that the WZ QCM acts efficiently for a larger value of Alice's state parameter $\alpha^2$. In this case, Eve manages to remain hidden in the QKD process in spite of generating the secret key,.

\subsubsection{Efficiency of modified BH QCM in successful QKD protocol}
% \usepackage{multirow}
% \usepackage{graphicx}
\begin{table}[h]
	\resizebox{\columnwidth}{!}{%
		\begin{tabular}{|c|c|c|c|c|}
			\hline
			\textbf{$\xi$} & \textbf{$\alpha^2$} & \textbf{$F_{{BH}}(\rho_E,\rho_E')$} & \textbf{$D(\rho_E,\rho_E')^2$} & \textbf{$D_{HS}^{{BH}}(\rho_{id},\rho_E)$}\\ \hline
			\multirow{5}{*}{0.1} & 0.241 & 0.8283 & [0,0.3139) & [0,0.6279) \\ \cline{2-5} 
			& 0.30 & 0.8976 & [0,0.1943) & [0,0.3886) \\ \cline{2-5} 
			& 0.35 & 0.9424 & [0,0.1118) & [0,0.2237) \\ \cline{2-5} 
			& 0.40 & 0.9744 & [0,0.0504) & [0,0.1009) \\ \cline{2-5} 
			& 0.445 & 0.9936 & [0,0.0154) & [0,0.0308) \\ \hline
			\multirow{7}{*}{0.2} & 0.155 & 0.8286 & [0,0.3134) & [0,0.6268) \\ \cline{2-5} 
			& 0.20 & 0.8704 & [0,0.2424) & [0,0.4848) \\ \cline{2-5} 
			& 0.25 & 0.9100 & [0,0.1719) & [0,0.3438) \\ \cline{2-5} 
			& 0.30 & 0.9424 & [0,0.1118) & [0,0.2237) \\ \cline{2-5} 
			& 0.35 & 0.9676 & [0,0.0637) & [0,0.1275) \\ \cline{2-5} 
			& 0.40 & 0.9856 & [0,0.0285) & [0,0.0571) \\ \cline{2-5} 
			& 0.426 & 0.9921 & [0,0.0157) & [0,0.0314) \\ \hline
			\multirow{9}{*}{0.3} & 0.001 & 0.8406 & [0,0.2933) & [0,0.5866) \\ \cline{2-5} 
			& 0.05 & 0.8704 & [0,0.2424) & [0,0.4848) \\ \cline{2-5} 
			& 0.10 & 0.8976 & [0,0.1943) & [0,0.3886) \\ \cline{2-5} 
			& 0.15 & 0.9216 & [0,0.1506) & [0,0.3013) \\ \cline{2-5} 
			& 0.20 & 0.9424 & [0,0.1118) & [0,0.2237) \\ \cline{2-5} 
			& 0.25 & 0.9600 & [0,0.0784) & [0,0.1568) \\ \cline{2-5} 
			& 0.30 & 0.9744 & [0,0.0505) & [0,0.1011) \\ \cline{2-5} 
			& 0.35 & 0.9856 & [0,0.0286) & [0,0.0572) \\ \cline{2-5} 
			& 0.39 & 0.9923 & [0,0.0154) & [0,0.0308) \\ \hline
			\multirow{7}{*}{0.4} & 0.001 & 0.9602 & [0,0.0781) & [0,0.1562) \\ \cline{2-5} 
			& 0.05 & 0.9676 & [0,0.0637) & [0,0.1275) \\ \cline{2-5} 
			& 0.10 & 0.9744 & [0,0.0505) & [0,0.1011) \\ \cline{2-5} 
			& 0.15 & 0.9804 & [0,0.0388) & [0,0.0776) \\ \cline{2-5} 
			& 0.20 & 0.9856 & [0,0.0286) & [0,0.0572) \\ \cline{2-5} 
			& 0.25 & 0.9900 & [0,0.0199) & [0,0.0398) \\ \cline{2-5} 
			& 0.28 & 0.9923 & [0,0.0154) & [0,0.0308) \\ \hline
			\multirow{2}{*}{0.455} & 0.001 & 0.9919 & [0,0.0161) & [0,0.0321) \\ \cline{2-5} 
			& 0.011 & 0.9923 & [0,0.0154) & [0,0.0308) \\ \hline
		\end{tabular}%
	}
	\caption{Analysis of modified BH QCM based on Eve's cloning machine's parameter $\xi$, Alice's secret state parameter $\alpha^2$ and, the HS distance between the input state $\rho_{id}$ and the output state $\rho_E$.}
	\label{table_{BH}_qcm}
\end{table}

When Eve uses the modified BH QCM to extract information from the intercepted state, it can be noted that the fidelity $F(\rho_E,\rho_E')$ of the modified BH QCM not only depends on Alice's state parameter $\alpha^2$ but also on the cloning machine parameter $\xi$. Therefore, from table \ref{table_{BH}_qcm} we can see that for a given value of $\alpha^2$, the upper bound of HS distance decreases, i.e., the efficiency of the cloning machine increases as $\xi$ increases. This means that for a given value of $\alpha^2$ and a larger value of $\xi$, Eve can copy the intercepted state in a better way. Therefore, though Eve can extract information using the QCM but its effect cannot be seen in the generation of secret key. Thus, Alice and Bob are able to generate the secret key as $R_{lb}>0$, in this case, in spite of Eve's intervention in the QKD protocol.
\section{Conclusion}
To summarize, we have analyzed the lower bound of key rate proposed by Woodhead \cite{woodhead_2013} for a variant of BB84 protocol. This lower bound does not assure the success of the protocol, i.e., successful key distribution between the communicating parties. This means that if the key rate is found to be negative, the protocol is aborted. To overcome this problem, we modified the lower bound given by Woodhead in such a way that the success of the protocol is assured. The secret key rate depends non-linearly on both error rate $\delta_z$ and Eve's fidelity $F(\rho_E,\rho_E')$. For a successful key distribution scheme, we find that for a given fidelity, $F(\rho_E,\rho_E')\in (0.8281,0.9922)$, there exists error rate $\delta_z\in (0,\delta_{z_1})$, such that $\delta_{z_1}\in (0,0.305)$. The modified lower bound depends only upon the overlapping between the density operators $\rho_E$ and $\rho_E'$, that is obtained by Eve after applying the cloning transformation on the intercepted state $\ket{\phi}$ and $\ket{\phi'}$ respectively, and it guaranteed a successful QKD scheme. In particular, we have studied two types of state dependent QCMs, namely, WZ QCM and modified BH QCM that may be used by the eavesdropper to extract the information from the intercepted state. The modified BH QCM can be constructed by setting the cloning machine parameters $\eta$ and $\xi$ as $\eta=0$ and $\xi\neq 0$. In both of these cases, we have shown that it is possible for Alice and Bob to distill a secret key in spite of the presence of the eavesdropper.	Moreover, we use the inequality between $F(\rho_E,\rho_E')$ and $D(\rho_E,\rho_E')$ to determine the range of $D(\rho_E,\rho_E')$, for a given value of $F(\rho_E,\rho_E')$. Since the Hilbert Schmidt distances $D_{HS}(\rho_{id},\rho_E)$ and $D_{HS}(\rho_{id}',\rho_E')$ represent the quality of the output of the cloning machine so we have estimated $D_{HS}(\rho_{id},\rho_E)$ and $D_{HS}(\rho_{id}',\rho_E')$ in terms of $D(\rho_E,\rho_E')$ and thereby discussed the effect of efficiency of the QCM on the generation of the secret key that is required for a successful key distribution protocol.
%For a larger value of Alice's state parameter $\alpha^2$, WZ QCM acts efficiently. In case of modified BH QCM, for a given value of $\alpha^2$, if the cloning machine parameter $\xi$ increases, the efficiency of the cloning machine increases. In both the cases, the eavesdropper can extract the information from the intercepted state, without revealing her presence in the QKD protocol. Thus, we have shown that there exist QCMs that the eavesdropper may use, and still generation of the secret key can be assured between the communicating parties with $R_{lb}>0$.\\
The use of asymmetric quantum cloning transformations in the proposed methodology is still unexplored, and we leave it as an open question how our results extend in this case.	
%Further, we analyze the symmetric WZ QCM. We find that the fidelity $F_{WZ}(\rho_E,\rho_E')$ is only dependent upon Alice's secret state parameter $\alpha^2$. As Alice increases the value of the secret state parameter $\alpha^2$, the fidelity $F(\rho_E,\rho_E')$ increases, correspondingly the range of error rate $\delta_z$ increases when Eve uses the WZ QCM. we observed that, Alice is restricted to choosing $\alpha^2\in (0.293,0.456)$ so that the secure key formation is ensured between Alice and Bob, even if the eavesdropper is present in the protocol. Next, we modify the BH QCM in such a way that eve's fidelity depends not only on Alice's secret state parameter $\alpha^2$, but also on the cloning machine parameter $\xi$. We find that for a given value of $\xi\in (0,0.455)$, there exist certain values of $\alpha^2\in (0,0.5)$ for which Eve manages to remain hidden in the key distribution process, and the key is distributed between the communicating parties successfully. We have shown that is possible for Alice and Bob to distill a secure key, in spite of the increasing interval of $\delta_z$.\\

%	\section{Data availability statement}
%	Data sharing not applicable to this article as no datasets were generated or analyzed during the current study.

\section{Declarations}
\subsection{Author contribution statement}
\textbf{RJ, SA}: Conceptualization, Methodology, Software, Writing- Original draft preparation, Visualization, Writing- Reviewing and Editing.

\subsection{Competing interests}
The authors have no competing interests to declare that are relevant to the content of this article.

\subsection{Data Availability Statement}
Data sharing is not applicable to this article as no data sets were generated or analyzed during the current study.
\balance

\appendix
\section{The relationship between the usual $Z$ and $X$ basis and new $Z'$ and $X'$ basis}
Consider the following unitary transformation $U'$ defined as
\begin{align}
	U'=\begin{pmatrix}
		\cos \theta	& -\sin \theta \\
		\sin \theta	&\cos \theta 
	\end{pmatrix}
\end{align}
The transformation $U'$ transforms $\ket{0}$ and $\ket{1}$ as follows
\begin{align}
	\begin{split}
		\ket{0}&\longrightarrow \cos \theta \ket{0}+\sin \theta \ket{1}\\
		\ket{1}&\longrightarrow -\sin \theta \ket{0}+\cos \theta \ket{1}
	\end{split}
\end{align}
This transforms the usual $Z$ basis, represented by $B_{usual}^{z}=\{(1,0),(0,1)\}\equiv \{a^{(0)},a^{(1)}\}$ to $Z'$ basis, given by
\begin{align}
	\begin{split}
		a^{\theta}&=\cos \theta a^{(0)}+\sin \theta a^{(1)}\\
		a^{\frac{\pi}{2}+\theta}&=-\sin \theta a^{(0)}+\cos \theta a^{(1)}\\
	\end{split}
\end{align} 
where $\cos \theta=\alpha$ and $\sin \theta=\beta$.	Similarly, the usual $X$ basis, represented by $B_{usual}^{x}=\{(\frac{1}{\sqrt{2}},\frac{1}{\sqrt{2}}),(\frac{1}{\sqrt{2}},-\frac{1}{\sqrt{2}})\}\equiv\{b^{(0)},b^{(1)}\}$ to $X'$ basis, given by
\begin{align}
	\begin{split}
		b^{\theta}&=(\cos \theta-\sin \theta) b^{(0)}+(\sin \theta+\cos \theta) b^{(1)}\\
		b^{\frac{\pi}{2}+\theta}&=(\cos \theta+\sin \theta) b^{(0)}+(\sin \theta-\cos \theta) b^{(1)}
	\end{split}
\end{align}
where $\cos \theta-\sin \theta=\gamma$ and $\sin \theta+\cos \theta=\delta$.

\section{Fidelity between $\rho$ and $\sigma$}
In this section, we will calculate the fidelity between two density matrices $\rho$ and $\sigma$.\\
The fidelity $F(\rho, \sigma)$ is given by \cite{muller_2023}
\begin{align}
	F(\rho,\sigma)=\left(\sum_{i}\sqrt{\lambda_i(\rho \sigma)}\right)^2
	\label{fid_muller}
\end{align}
where $\lambda_i$ is the $i^{th}$ eigenvalue of the product $\rho \sigma$. 

\subsection{Fidelity of WZ QCM for $\ket{\phi}$ and $\ket{\phi'}$}
The fidelity between the density matrices $\rho_E$ and $\rho_E'$ given in (\ref{wz_rhoe}-\ref{wz_rhoe'}) maybe calculated as
\begin{align}
	\rho_E \rho_E'= \begin{bmatrix}
		\alpha^2 \beta^2 &  0 \\
		0 & \alpha^2 \beta^2
	\end{bmatrix}
\end{align} 
The eigenvalues of $\rho_E \rho_E'$ are given by $\alpha^2 \beta^2, \alpha^2 \beta^2$. Therefore, substituting these values in (\ref{fid_muller}), we get
\begin{align}
	F_{WZ}(\rho_E,\rho_E')&=\left(\sqrt{\alpha^2 \beta^2}+\sqrt{\alpha^2 \beta^2}\right)^2\notag \\
	&=\left(2 \sqrt{\alpha^2 \beta^2}\right)^2\notag \\
	&=4\alpha^2 \beta^2
	\label{wz_fid_a1}
\end{align}

\subsection{Fidelity of the modified BH QCM}
The eigenvalues of the product of the density matrices $\rho_E$ and  $\rho_E'$ obtained by Eve after applying the cloning transformation (\ref{bh_qcm}) and tracing out Bob's bits, can be given by $\{(\alpha^2 +\xi (\beta^2-\alpha^2))(\beta^2+\xi (\alpha^2-\beta^2)), (\alpha^2 +\xi (\beta^2-\alpha^2))(\beta^2+\xi (\alpha^2-\beta^2))\}$. Therefore, the fidelity $F_{{BH}}(\rho_E,\rho_E')$ is given by
\begin{align}
	\begin{split}
		F_{{BH}}(\rho_E,\rho_E')&=\bigg(\sqrt{(\alpha^2 +\xi (\beta^2-\alpha^2))(\beta^2+\xi (\alpha^2-\beta^2))}\\
		&+\sqrt{(\alpha^2 +\xi (\beta^2-\alpha^2))(\beta^2+\xi (\alpha^2-\beta^2))}\bigg)^2\\
		&=4(\alpha^2 +\xi (\beta^2-\alpha^2))(\beta^2+\xi (\alpha^2-\beta^2))
	\end{split}
\end{align}
Now using the normalization condition $\beta^2=1-\alpha^2$, we obtain
\begin{align}
	F_{{BH}}(\rho_E,\rho_E')=4\left(\alpha^2(1-\alpha^2)(1-2\xi)^2+\xi(1-\xi)\right)
	\label{fid_our_qcm1_a1}
\end{align}


\begin{thebibliography}{9}
	%
	\bibitem{merkle_1978} R.C. Merkle, Comm. of the ACM, \textbf{21}, 294 (1978).
	\bibitem{gisin_2002} N. Gisin, G. Ribordy, Q. Tittel, and H. Zbinden, Rev. of Mod. Phys. \textbf{74}, 145 (2002).
	\bibitem{ballentine_2014} L.E. Ballentine L.E., World Scientific Publishing Company (2014).
	\bibitem{bb84} C. H. Bennett, and G. Brassard, In: Proceedings of the IEEE International Conference on Computers, Systems, and Signal Processing, Bangalore, India (IEEE, New York, 1984), pp. 175-179.
	\bibitem{jain_2024} R. Jain, and S. Adhikari, Eu. Phys. J. D \textbf{78}, 145 (2024).
	\bibitem{scarani_2009} V. Scarani, H. Bechmann-Pasquinucci, N.J. Cerf et al., Rev. Mod. Phys. \textbf{81}, 1301 (2009).
	\bibitem{tan_2025} H. Tan, M. Petrov, W. Zhang et al., PRX Quantum \textbf{6}, 040331 (2025).
	\bibitem{ekert_1991} A. K. Ekert, Phys. Rev. Lett. \textbf{67}, 661 (1991).
	\bibitem{fung_2006} C.F. Fung, K. Tamaki, and H.K. Lo, Phys. Rev. A \textbf{73}, 012337 (2006).
	\bibitem{young_2026} M. Young, M. Lucamarini, and S. Pirandola, Sci. Rep. \textbf{16}, 598 (2026).
	\bibitem{metger_2023} T. Metger and R. Renner, Nat. Comm. \textbf{14}, 5272 (2023).
	\bibitem{dixon_2014} A. R. Dixon, and H. Sato, Sci. Rep. \textbf{4}, 7275 (2014). 
	\bibitem{pastushenko_2023} V. A. Pastushenko, and D. A. Kronberg, Entropy \textbf{25}, 956 (2023). 
	\bibitem{horvath_2011} T. Horvath, L. B. Kish, and J. Scheuer, Europhy. Lett. \textbf{94} 28002, (2011).
	\bibitem{tang_2019} B. Y. Tang, B. Liu, Y. P. Zhai, C. Q. Wu, and W. R. Yu, Sci. Rep. \textbf{9}, 15733 (2019).
	\bibitem{pirandola_2020} S. Pirandola, et al., Adv. Opt. Photonics \textbf{12}, 1012 (2020).
	\bibitem{barrett_2005} J. Barrett, L. Hardy and A. Kent, Phys. Rev. Lett. \textbf{95}, 010503 (2005).
	\bibitem{biham_2002} E. Biham, M. Boyer,G. Brassard, J. V. Graaf, and T. Mor, Algorithmica, \textbf{34}, 372 (2002). 
	\bibitem{acin_2007} A. Acin, N. Brunner, N. Gisin et al., Phys. Rev. Lett. \textbf{98}, 230501 (2007). 
	\bibitem{renner_thesis} R. Renner, Arxiv preprint arXiv:0512258 (2005).
	\bibitem{diamanti_2016} E. Diamanti, H.K. Lo, B. Qi et al. npj Quant. Inf. \textbf{2}, 16025 (2016).
	\bibitem{gottesman_2004}  D. Gottesman, H.-K. Lo, N. L¨utkenhaus, and J. Preskill, Quant. Inf. Comp. \textbf{4}, 325 (2004).
	\bibitem{liu_2025} H.D. Liu, B. Tian, Y.Q. Chen et al., Nonlinear Dyn. \textbf{113}, 3655 (2025).
	\bibitem{shan_2024} H.W. Shan, B. Tian, C.D. Chen et al., Qual. Theory Dyn. Syst. \textbf{23}, 267 (2024). 
	\bibitem{feng_2025} C.H. Feng, B. Tian, and X.Y. Gao, Qual. Theory Dyn. Syst. \textbf{24}, 100 (2025).
	\bibitem{wang_2026} G. Wang, Z. Tan, X.Y. Gao and J.G. Liu, Appl. Math. Lett. \textbf{172}, 109720 (2026).
	\bibitem{gao1_2024} X.Y. Gao, Appl. Math. Lett. \textbf{152}, 109018 (2024).
	\bibitem{gao1_2025} X.Y. Gao, China Ocean Eng. \textbf{39}, 541 (2025).
	\bibitem{gao2_2024} X.Y. Gao, Chin. J. Phys. \textbf{92}, 1233 (2024).
	\bibitem{gao2_2025} X.Y. Gao, J.G. Liu and G.W. Wang, Appl. Math. Lett. \textbf{171}, 109615 (2025).
	\bibitem{devetak_2005} I. Devetak and A. Winter, Proc. R. Soc. A \textbf{461}, 207 (2005).
	\bibitem{kraus_2005} B. Kraus, N. Gisin, and R. Renner, Phys. Rev. Lett. \textbf{95}, 080501 (2005).
	\bibitem{maroy_2010} Ø. Marøy, L. Lydersen, and J. Skaar, Phys.Rev.A \textbf{82}, 032337 (2010).
	\bibitem{koashi_2009} M. Koashi, New J. Phys. \textbf{11}, 045018 (2009).
	\bibitem{curty_2009} M. Curty, T. Moroder, X. Ma, H. K. Lo, and N. Lutkenhaus, Phys. Rev. A \textbf{79}, 032335 (2009).
	\bibitem{kaur_2022} E. Kaur, K. Horodecki, and S. Das, Phys. Rev. App. \textbf{18}, 054033 (2022). 
	\bibitem{zhang_2017} Z. Zhang, Q. Zhao, M. Razavi, and X. Ma, Phys. Rev. A \textbf{95}, 012333 (2017).
	\bibitem{woodhead_2013} E. Woodhead, Phys. Rev. A \textbf{88}, 012331 (2013).
	\bibitem{shruti_2015} S. Dogra, K. Dorai, and Arvind, Phys. Rev. A. \textbf{91}, 022312 (2015).
	%\bibitem{gisin_1997} N. Gisin, and B. Huttner, Phys. Lett. A \textbf{228}, 13 (1997).
	\bibitem{wang_2023} P.Wang, Y. Zhang, Z. Lu, X. Wang and Y. Li, New J. Phys. \textbf{25}, 023019 (2023).
	\bibitem{sohr_2024} P.Sohr, S. Ecker, L. Bulla, M. Bohmann and R. Ursin, Phy. Rev. App. \text{22}, 024059 (2024).
	\bibitem{shor_2000} P.W. Shor and J. perskill, Phys. Rev. Lett. \textbf{85}, 441 (2000).
	\bibitem{cover_1999_book} T. M. Cover, Elements of information theory. John Wiley \& Sons (1999).
	%\bibitem{neilsen_chuang} M. A. Nielsen, and I. L. Chuang, Quantum Computation and Quantum Information. Cambridge University Press, New Delhi (2008).
	\bibitem{gao_2024} X.Y. Gao, Qual. Theory Dyn. Syst. \textbf{23}, 202 (2024). 
	\bibitem{gao_2025} X.Y. Gao, Appl. Math. Lett. \textbf{159}, 109262 (2025).
	\bibitem{wooters_1982} W. K. Wooters, and W. H. Zurek, Nature London, \textbf{299}, 802, (1982).		
	\bibitem{muller_2023} A. Muller, arXiv:2309.10565v4, (2023).
	\bibitem{buzek_1996} V. Bužek, and M. Hillery, Phys. Rev. A \textbf{54}, 1844 (1996).
	%\bibitem{mceliece_2002_book} R. McEliece, The theory of information and coding, Cambridge University Press (2002).
	%\bibitem{wootters_1982}  W.K. Wootters, \& W.H. Zurek, Nature, \textbf{299}, 802 (1982).
	\bibitem{dononov_2000} V.V. Dodonov, O.V. Man'ko, V.I. Man'ko \& A. Wünsche, J. Mod. Opt. \textbf{47}, 633 (2000).
	\bibitem{witte_1999} C. Witte, and M. Trucks,	Phys. Lett. A \textbf{257}, 14 (1999).
	\bibitem{ozawa_2000} M. Ozawa, Entanglement measures and the Hilbert-Schmidt distance, Phys. Lett. A \textbf{268}, 158	(2000).
	\bibitem{jain_2025} R. Jain, and S. Adhikari, Phys. Scr. \textbf{100}, 035113 (2025).
	\bibitem{cincio_2018}  L. Cincio, Y. Subasi, A. T. Sornborger, and P. J. Coles, New J. Phys. \textbf{20}, 113022 (2018).
	\bibitem{li_2003}  J. Lee, M. S. Kim, and C. Brukner, Phys. Rev. Lett. \textbf{91}, 087902 (2003).
	\bibitem{neilsen_chuang} M. A. Nielsen, and I. L. Chuang, Quantum Computation and Quantum Information. Cambridge University Press, New Delhi (2008).
	\bibitem{coles_2019} P. Coles, M. Cerezo, and L. Cincio, Phys. Rev. A \textbf{100}, 022103 (2019).
	
\end{thebibliography}
\end{document}